\documentclass[11pt,a4paper]{article}
\usepackage{jheppub}
\usepackage{bbold} 
\usepackage{axodraw2}
\allowdisplaybreaks[1]

\newcommand{\la}[1]{\label{#1}}
\newcommand{\ba}{\begin{eqnarray}}
\newcommand{\ea}{\end{eqnarray}}
\newcommand{\rmi}[1]{{\mbox{\scriptsize #1}}}
\newcommand{\se}{section~}
\newcommand{\app}{appendix~}
\newcommand{\fig}{figure~}
\newcommand{\eq}{eq.~}
\newcommand{\eqs}{eqs.~}
\newcommand{\nr}[1]{(\ref{#1})}
\newcommand{\Tint}[1]{{\hbox{$\sum$}\!\!\!\!\!\!\!\int\,}_{\!\!\!\!\!\!\raise-0.2ex\hbox{$\scriptstyle{#1}$}}}
\newcommand{\ep}{\varepsilon}
\newcommand{\ccdot}{\!\cdot\!}
\newcommand{\secs}{sections~}
\newcommand{\figs}{figures~}
\newcommand{\setN}{\mathbb{N}}
\newcommand{\setZ}{\mathbb{Z}}
\newcommand{\rmii}[1]{{\mbox{\tiny\rm{#1}}}}
\newcommand{\Tinti}[1]{{{\Sigma}\!\!\!\!\raise0.3ex\hbox{$\int$}_\rmii{${#1}$}}}
\newcommand{\sumintL}{L}
\newcommand{\nabc}{{\nu_1,\nu_2,\nu_3}}
\newcommand{\eabc}{{\eta_1,\eta_2,\eta_3}}
\newcommand{\ci}{{\cal I}}
\newcommand{\Nc}{N_\rmi{c}}
\newcommand{\Nf}{N_\rmi{f}}
\renewcommand{\vec}[1]{{\bf #1}}
\newcommand{\gb}{g_\rmi{B}}

\newcommand{\pic}[1]{\;\parbox[c]{30pt}{\begin{picture}(30,30)(0,0)
\SetWidth{1.0}\SetScale{1.0} #1 \end{picture}}\;}
\newcommand{\picb}[1]{\;\parbox[c]{45pt}{\begin{picture}(45,30)(0,0)
\SetWidth{1.0}\SetScale{1.0} #1 \end{picture}}\;}
\newcommand{\picc}[1]{\;\parbox[c]{60pt}{\begin{picture}(60,30)(0,0)
\SetWidth{1.0}\SetScale{1.0} #1 \end{picture}}\;}
\def\Lwidth{1}

\def\Agl(#1,#2)(#3,#4,#5){\PhotonArc(#1,#2)(#3,#4,#5){\Lwidth}
{6.283 #3 mul 360 div #4 #5 sub #4 #5 sub mul sqrt mul Ldensity mul}}
\def\Lgl(#1,#2)(#3,#4){\Photon(#1,#2)(#3,#4){\Lwidth}
{#1 #3 sub #1 #3 sub mul #2 #4 sub #2 #4 sub mul add sqrt Ldensity mul}}
\def\Agh(#1,#2)(#3,#4,#5){\DashArrowArc(#1,#2)(#3,#4,#5){1}}
\def\Ahg(#1,#2)(#3,#4,#5){\DashArrowArcn(#1,#2)(#3,#5,#4){1}}
\def\Lgh(#1,#2)(#3,#4){\DashArrowLine(#1,#2)(#3,#4){1}}
\def\Lhg(#1,#2)(#3,#4){\DashArrowLine(#3,#4)(#1,#2){1}}
\def\Ahh(#1,#2)(#3,#4,#5){\DashCArc(#1,#2)(#3,#4,#5){1}}
\def\Lhh(#1,#2)(#3,#4){\DashLine(#1,#2)(#3,#4){1}}
\def\Aqu(#1,#2)(#3,#4,#5){\ArrowArc(#1,#2)(#3,#4,#5)}
\def\Auq(#1,#2)(#3,#4,#5){\ArrowArcn(#1,#2)(#3,#5,#4)}
\def\Lqu(#1,#2)(#3,#4){\ArrowLine(#1,#2)(#3,#4)}
\def\Luq(#1,#2)(#3,#4){\ArrowLine(#3,#4)(#1,#2)}
\def\Aqq(#1,#2)(#3,#4,#5){\CArc(#1,#2)(#3,#4,#5)}
\def\Lqq(#1,#2)(#3,#4){\Line(#1,#2)(#3,#4)}
\def\Asc(#1,#2)(#3,#4,#5){\CArc(#1,#2)(#3,#4,#5)}
\def\Lsc(#1,#2)(#3,#4){\Line(#1,#2)(#3,#4)}
\def\TopoVRoi(#1){\;\pic{#1(15,15)(15,0,180) #1(15,15)(15,180,360)%
 \GCirc(0,15){3}{0} \GBoxc(30,15)(6,6){0}}\;}
\def\TopoVRoo(#1){\;\pic{#1(15,15)(15,0,180) #1(15,15)(15,180,360)%
 \GCirc(0,15){3}{0} \GCirc(30,15){3}{0}}\;}
\def\TopoVRooo(#1){\pic{#1(15,15)(15,-30,90) #1(15,15)(15,90,210)%
 #1(15,15)(15,210,330) \GCirc(15,30){3}{0} \GCirc(2,7.5){3}{0}%
 \GCirc(28,7.5){3}{0}}}
\def\ToptVS(#1,#2,#3){\pic{#1(15,15)(15,0,180) #2(15,15)(15,180,360)%
 #3(30,15)(0,15)}}
\def\ToptVE(#1,#2){\picc{#1(15,15)(15,0,360) #2(45,15)(15,-180,180)}}
\def\ToprVM(#1,#2,#3,#4,#5,#6){\pic{#3(15,15)(15,-30,90) #1(15,15)(15,90,210)%
 #2(15,15)(15,210,330) #5(15,15)(2,7.5) #6(15,30)(15,15) #4(28,7.5)(15,15)}}
\def\ToprVV(#1,#2,#3,#4,#5){\!\!\picb{#2(26.25,15)(15,256,76)%
 #3(30,30)(15,30) #1(18.75,15)(15,104,284) #4(15,30)(22.5,0)%
 #5(30,30)(22.5,0)}\!\!}
\def\ToprVB(#1,#2,#3,#4){\picb{#1(30,15)(15,-120,120) #2(30,15)(15,120,240)%
 #3(15,15)(15,60,300) #4(15,15)(15,-60,60)}}
\def\ToprVBB(#1,#2,#3,#4,#5,#6){\picb{#1(30,15)(15,-90,90)%
 #2(15,30)(30,30) #3(15,15)(15,90,270) #4(30,0)(15,0)%
 #5(30,30)(30,0) #6(15,0)(15,30)}}
\def\ToprVBT(#1,#2,#3,#4,#5){\picc{#3(15,15)(15,0,360) #5(45,30)(45,0)
 #1(45,15)(15,-90,90) #2(45,15)(15,90,180) #4(45,15)(15,180,270)}}
\def\ToprVTT(#1,#2,#3,#4){\picc{#2(30,15)(15,0,180) #4(30,15)(15,180,360)%
 #1(52.5,15)(7.5,-180,180) #3(7.5,15)(7.5,0,360)}}
\def\TopfVX(#1,#2,#3,#4,#5,#6,#7,#8,#9){\picb{#1(15,15)(15,90,270)%
 #2(30,15)(15,-90,90) #4(30,30)(15,30) #3(15,0)(30,0) #6(15,0)(15,15)%
 #5(15,15)(30,30) #8(15,30)(20,25) #8(25,20)(30,15) #7(30,15)(30,0)%
 #9(15,15)(30,15)}}
\def\TopfVH(#1,#2,#3,#4,#5,#6,#7,#8,#9){\picb{#1(15,15)(15,90,270)%
 #2(30,15)(15,-90,90) #4(30,30)(15,30) #3(15,0)(30,0) #6(15,0)(15,15)%
 #5(15,15)(15,30) #8(30,30)(30,15) #7(30,15)(30,0) #9(15,15)(30,15)}}
\def\TopfVW(#1,#2,#3,#4,#5,#6,#7,#8){\pic{#1(15,15)(15,90,180)%
 #3(15,15)(15,180,270) #2(15,15)(15,270,360) #4(15,15)(15,0,90)%
 #5(15,15)(15,30) #7(15,15)(15,0) #6(0,15)(15,15) #8(30,15)(15,15)}}
\def\TopfVV(#1,#2,#3,#4,#5,#6,#7,#8){\!\!\picb{#2(26.25,15)(15,256,346)%
 #3(26.25,15)(15,-14,76) #4(30,30)(15,30) #1(18.75,15)(15,104,284)%
 #7(22.5,0)(15,30) #6(30,30)(26.25,15) #8(26.25,15)(22.5,0)%
 #5(25.25,15)(39.8,11.4)}\!\!}
\def\TopfVB(#1,#2,#3,#4,#5,#6,#7){\picb{#2(30,15)(15,-120,120)%
 #6(30,15)(15,120,180) #5(30,15)(15,180,240) #1(15,15)(15,60,300)%
 #4(15,15)(15,-60,0) #3(15,15)(15,0,60) #7(30,15)(15,15)}}
\def\TopfVN(#1,#2,#3,#4,#5,#6,#7){\picb{#1(15,15)(15,90,270)%
 #2(30,15)(15,-90,90) #4(30,30)(15,30) #3(15,0)(30,0)%
 #5(15,0)(15,30) #6(30,30)(30,0) #7(15,30)(30,0)}}
\def\TopfVU(#1,#2,#3,#4,#5,#6,#7){\pic{#3(15,15)(15,0,90)%
 #2(15,15)(15,90,180) #4(15,15)(15,180,270) #1(15,15)(15,270,360)%
 #6(0,15)(15,30) #7(15,0)(0,15) #5(30,15)(15,0)}}
\def\TopfVT(#1,#2,#3,#4,#5,#6){\pic{#1(15,15)(15,90,210)%
 #2(15,15)(15,210,330) #3(15,15)(15,-30,90) #4(2,7.5)(15,30)%
 #6(28,7.5)(2,7.5) #5(15,30)(28,7.5)}}
\def\TopfVMT(#1,#2,#3,#4,#5,#6,#7,#8){\picb{%
#1(15,15)(15,0,90)%
#2(15,15)(15,90,180)%
#3(15,15)(15,180,270)%
#4(15,15)(15,270,360)%
#5(15,30)(15,15)%
#6(0,15)(15,15)%
#7(15,15)(15,0)%
#8(37.5,15)(7.5,-180,180)}}
\def\TopfVMB(#1,#2,#3,#4,#5,#6,#7,#8,#9){\picb{%
#1(15,30)(30,30)%
#2(15,15)(15,90,180)%
#3(15,15)(15,180,270)%
#4(15,0)(30,0)%
#5(15,30)(15,15)%
#6(0,15)(15,15)%
#7(15,15)(15,0)%
#8(30,30)(30,0)%
#9(30,15)(15,-90,90)}}
\def\TopfVVTa(#1,#2,#3,#4,#5,#6,#7){\picb{%
#1(15,15)(15,0,60)%
#2(15,15)(15,60,180)%
#3(15,15)(15,180,300)%
#4(15,15)(15,300,360)%
#5(0,15)(22.5,28)%
#6(0,15)(22.5,2)%
#7(37.5,15)(7.5,-180,180)}}
\def\TopfVVBa(#1,#2,#3,#4,#5,#6,#7,#8){\picb{%
#1(15,30)(30,30)%
#2(15,15)(15,90,180)%
#3(15,15)(15,180,270)%
#4(15,0)(30,0)%
#5(0,30)(15,-90,0)%
#6(0,0)(15,0,90)%
#7(30,30)(30,0)%
#8(30,15)(15,-90,90)}}
\def\TopfVVTb(#1,#2,#3,#4,#5,#6,#7){\!\!\picc{%
#1(26.25,15)(15,-104,-5)%
#4(26.25,15)(15,-5,76)%
#2(15,30)(30,30)%
#3(18.75,15)(15,104,284)%
#5(15,30)(22.5,0)%
#6(30,30)(22.5,0)%
#7(49,12)(7.5,-180,180)}\!\!}
\def\TopfVVBb(#1,#2,#3,#4,#5,#6,#7,#8){\picc{%
#1(30,30)(45,30)%
#4(22.5,0)(45,0)%
#2(15,30)(30,30)%
#3(18.75,15)(15,104,284)%
#5(15,30)(22.5,0)%
#6(30,30)(22.5,0)%
#7(45,30)(45,0)%
#8(45,15)(15,-90,90)}}
\def\TopfVBT(#1,#2,#3,#4,#5,#6){\picc{%
#1(15,15)(15,60,300)%
#2(30,15)(15,120,240)%
#3(15,15)(15,-60,60)%
#4(30,15)(15,0,120)%
#5(30,15)(15,-120,0)%
#6(52.5,15)(7.5,-180,180)}}
\def\TopfVBBa(#1,#2,#3,#4,#5,#6,#7){\picc{%
#1(15,15)(15,60,300)%
#2(30,15)(15,120,240)%
#3(15,15)(15,-60,60)%
#4(30,15)(15,30,120)%
#5(30,15)(15,-120,-30)%
#6(30,15)(15,-30,30)%
#7(45,15)(7.5,-100,100)}}
\def\TopfVBSB(#1,#2,#3,#4,#5,#6,#7,#8,#9){\picc{%
#1(15,15)(15,90,270)%
#2(15,30)(15,0)%
#3(30,30)(30,0)%
#4(45,30)(45,0)%
#5(45,15)(15,-90,90)%
#6(15,30)(30,30)%
#7(15,0)(30,0)%
#8(30,30)(45,30)%
#9(30,0)(45,0)}}
\def\TopfVBST(#1,#2,#3,#4,#5,#6,#7,#8){\picc{%
#1(15,15)(15,90,270)%
#2(15,30)(15,0)%
#3(30,30)(30,0)%
#4(15,30)(30,30)%
#5(15,0)(30,0)%
#6(30,15)(15,0,90)%
#7(30,15)(15,-90,0)%
#8(52.5,15)(7.5,-180,180)}}
\def\TopfVSS(#1,#2,#3,#4,#5,#6,#7,#8){\picc{%
#1(15,15)(15,90,270)%
#2(15,30)(15,0)%
#3(15,15)(15,0,90)%
#4(15,15)(15,-90,0)%
#5(45,15)(15,90,180)%
#6(45,15)(15,180,270)%
#7(45,30)(45,0)%
#8(45,15)(15,-90,90)}}
\def\TopfVSTT(#1,#2,#3,#4,#5,#6,#7){\picc{%
#1(15,15)(15,90,270)%
#2(15,30)(15,0)%
#3(15,15)(15,0,90)%
#4(15,15)(15,-90,0)%
#5(37.5,15)(7.5,0,180)%
#6(37.5,15)(7.5,180,360)%
#7(52.5,15)(7.5,-180,180)}}
\def\TopfVTST(#1,#2,#3,#4,#5,#6,#7){\picc{%
#1(7.5,15)(7.5,0,360)%
#2(30,15)(15,90,180)%
#3(30,15)(15,180,270)%
#4(30,30)(30,0)%
#5(30,15)(15,0,90)%
#6(30,15)(15,-90,0)%
#7(52.5,15)(7.5,-180,180)}}
\def\TopfVTTTT(#1,#2,#3,#4,#5,#6){\picc{%
#1(7.5,15)(7.5,0,360)%
#2(22.5,15)(7.5,0,180)%
#3(22.5,15)(7.5,180,360)%
#4(37.5,15)(7.5,0,180)%
#5(37.5,15)(7.5,180,360)%
#6(52.5,15)(7.5,-180,180)}}
\def\TopfVBBB(#1,#2,#3,#4,#5,#6,#7,#8,#9){\picb{%
#8(20,20)(20,0,60)%
#1(20,20)(20,60,120)%
#4(20,20)(20,120,180)%
#2(20,20)(20,180,240)%
#6(20,20)(20,240,300)%
#3(20,20)(20,300,360)%
#7(20,0)(10,15,165)%
#5(2.7,30)(10,-100,50)%
#9(37.3,30)(10,130,280)}}
\def\TopfVBBT(#1,#2,#3,#4,#5,#6,#7,#8){\picb{%
#1(15,15)(15,0,60)%
#4(15,15)(15,60,120)%
#2(15,15)(15,120,240)%
#6(15,15)(15,240,300)%
#3(15,15)(15,300,360)%
#5(15,30)(10,195,345)%
#7(15,0)(10,15,165)%
#8(37.5,15)(7.5,-180,180)}}
\def\TopfVBTT(#1,#2,#3,#4,#5,#6,#7){\picb{%
#1(20,15)(15,45,135)%
#2(20,15)(15,135,240)%
#5(20,15)(15,240,300)%
#3(20,15)(15,-60,45)%
#6(20,0)(10,15,165)%
#4(5,30)(5.5,-45,315)%
#7(35,30)(5.5,-135,225)}}
\def\TopfVTTT(#1,#2,#3,#4,#5,#6){\picb{%
#1(20,15)(15,45,135)%
#2(20,15)(15,135,270)%
#3(20,15)(15,-90,45)%
#5(20,-5.5)(5.5,-270,90)%
#4(5,30)(5.5,-45,315)%
#6(35,30)(5.5,-135,225)}}
\def\TopfVBBb(#1,#2,#3,#4,#5,#6,#7,#8,#9){\pic{%
#8(15,15)(15,0,60)%
#1(15,15)(15,60,120)%
#4(15,15)(15,120,180)%
#2(15,15)(15,180,240)%
#6(15,15)(15,240,300)%
#3(15,15)(15,300,360)%
#7(15,0)(7.5,15,165)%
#5(2,22.5)(7.5,-100,50)%
#9(28,22.5)(7.5,130,280)}}
\newcommand{\defDiag}[2]{\expandafter\newcommand%
  \csname diag-#1\endcsname{#2}}
\newcommand{\diag}[1]{\csname diag-#1\endcsname}
\newcommand{\TLfig}[1]{{\begin{array}{c}\diag{#1}\\[2ex] \mbox{\footnotesize #1}\end{array}}}
\newcommand{\picj}[1]{\;\parbox[c]{40pt}{\begin{picture}(40,40)(0,0)
\SetWidth{1.0}\SetScale{1.0} #1 \end{picture}}\;}
\newcommand{\picbj}[1]{\;\parbox[c]{60pt}{\begin{picture}(60,40)(0,0)
\SetWidth{1.0}\SetScale{1.0} #1 \end{picture}}\;}
\newcommand{\piccj}[1]{\;\parbox[c]{80pt}{\begin{picture}(80,40)(0,0)
\SetWidth{1.0}\SetScale{1.0} #1 \end{picture}}\;}
\newcommand{\sbx}{\scalebox{0.725}}
\defDiag{960}{\sbx{\picj{
\Asc(6.7,6.7)(9.3,135,315)
\Asc(6.7,33.3)(9.3,45,225)
\Asc(33.3,6.7)(9.3,225,45)
\Asc(33.3,33.3)(9.3,315,135)
\Lsc(13.3,0)(26.7,40)
\Lsc(0,13.3)(40,26.7)
\Lsc(0,26.7)(40,13.3)
\Lsc(13.3,40)(26.7,0)}}}
\defDiag{992}{\sbx{\picbj{
\Asc(20,20)(20,0,360)
\Lsc(0,20)(40,20)
\Asc(53.3,6.7)(9.3,225,45)
\Asc(53.3,33.3)(9.3,315,135)
\Lsc(40,20)(46.7,40)
\Lsc(40,20)(60,26.7)
\Lsc(40,20)(60,13.3)
\Lsc(40,20)(46.7,0)}}}
\defDiag{961}{\sbx{\picbj{
\Asc(20,20)(20,0,360)
\Asc(50,20)(10,0,360)
\Asc(20,2)(27,42,138)
\Asc(20,38)(27,222,318) }}}
\defDiag{841}{\sbx{\picj{
\Asc(20,20)(20,0,180)
\Asc(20,30)(22.36,-153.43,-26.57)
\Lsc(0,20)(40,20)
\Asc(20,10)(22.36,26.57,153.43)
\Asc(20,20)(20,-180,0)}}}
\defDiag{1008}{\sbx{\picbj{
\Asc(20,20)(20,0,360)
\Asc(50,20)(10,0,360)
\Lsc(40,20)(4,8)
\Lsc(40,20)(4,32)}}}
\defDiag{993}{\sbx{\!\!\picbj{
\Asc(35,20)(20,256,346)
\Asc(35,20)(20,-14,76)
\Lsc(40,39.5)(20,39.5)
\Asc(25,20)(20,104,284)
\Lsc(30,0)(20,39.5)
\Lsc(40,39.5)(30,0)
\Asc(38,23)(24,136,250)}\!\!}}
\defDiag{978}{\sbx{\piccj{
\Asc(20,20)(20,0,360)
\Asc(60,20)(20,0,360)
\Lsc(0,20)(80,20)}}}
\defDiag{952}{\sbx{\picj{
\Asc(20,20)(20,90,210)
\Asc(20,20)(20,210,330)
\Asc(20,20)(20,-30,90)
\Lsc(3,10)(20,40)
\Lsc(37,10)(3,10)
\Lsc(20,40)(37,10)}}}
\defDiag{1016}{\sbx{\picj{
\Asc(20,20)(20,0,90)
\Asc(20,20)(20,-180.,-90)
\Asc(40,0)(20,90,180)
\Asc(0,40)(20,-90,0)
\Asc(0,0)(20,0,90)
\Asc(20,20)(20,-90.,0)
\Asc(20,20)(20,90,180)}}}
\defDiag{1012}{\sbx{\picbj{
\Asc(20,20)(20,0,360)
\Asc(50,20)(10,0,360)
\Lsc(40,20)(20,20)
\Lsc(20,0)(20,40)}}}
\defDiag{1010}{\sbx{\picbj{
\Asc(20,20)(20,90,270)
\Asc(40,20)(20,-90,90)
\Lsc(40,40)(20,40)
\Lsc(20,0)(40,0)
\Lsc(20,0)(20,40)
\Lsc(40,40)(40,0)
\Lsc(20,40)(40,0)}}}
\defDiag{1009}{\sbx{\picbj{
\Asc(40,20)(20,-120,120)
\Asc(40,20)(20,120,180)
\Asc(40,20)(20,180,240)
\Asc(20,20)(20,60,300)
\Asc(20,20)(20,-60,0)
\Asc(20,20)(20,0,60)
\Lsc(40,20)(20,20)}}}
\defDiag{1020}{\sbx{\!\!\picbj{
\Asc(35,20)(20,256,346)
\Asc(35,20)(20,-14,76)
\Lsc(40,39.5)(20,39.5)
\Asc(25,20)(20,104,284)
\Lsc(30,0)(20,39.5)
\Lsc(40,39.5)(35,20)
\Lsc(35,20)(30,0)
\Lsc(34,20)(53,15)}\!\!}}
\defDiag{1011}{\sbx{\picj{
\Asc(20,20)(20,90,180)
\Asc(20,20)(20,180,270)
\Asc(20,20)(20,270,360)
\Asc(20,20)(20,0,90)
\Lsc(20,20)(20,40)
\Lsc(20,20)(20,0)
\Lsc(0,20)(20,20)
\Lsc(40,20)(20,20)}}}
\defDiag{1022}{\sbx{\picbj{
\Lsc(20,0)(40,0)
\Lsc(20,40)(40,40)
\Asc(20,20)(20,-180,-90)
\Asc(40,20)(20,-90,90)
\Lsc(20,20)(40,20)
\Lsc(20,0)(20,20)
\Lsc(40,20)(40,0)
\Lsc(20,20)(20,40)
\Lsc(40,20)(40,40)
\Asc(20,20)(20,90,180)}}}
\defDiag{511}{\sbx{\picbj{
\Asc(20,20)(20,90,270)
\Asc(40,20)(20,-90,90)
\Lsc(40,40)(20,40)
\Lsc(20,0)(40,0)
\Lsc(20,0)(20,20)
\Lsc(20,20)(40,40)
\Lsc(20,40)(28,33)
\Lsc(33,28)(40,20)
\Lsc(40,20)(40,0)
\Lsc(20,20)(40,20)}}}

\newcommand{\figureA}{%
\newcommand{\sbw}{\scalebox{0.9}}
\begin{figure}[tbp]
\begin{eqnarray*}
\sbw{ \pic{\Agl(15,15)(15,-90,270)} }
 -\sbw{ \pic{\Ahg(15,15)(15,-90,270)} }
 -\sbw{ \pic{\Auq(15,15)(15,-90,270)} }
\;\;;\quad
 {1\over8} \sbw{ \ToptVE(\Agl,\Agl) }
 +{1\over12} \sbw{ \ToptVS(\Agl,\Agl,\Lgl) }
 -{1\over2} \sbw{ \ToptVS(\Agl,\Ahg,\Lhg) }
 -{1\over2} \sbw{ \ToptVS(\Agl,\Auq,\Luq) }
\end{eqnarray*}
\vspace*{-6mm} 
\caption{\la{fig:12loop} Left panel: Feynman diagrams contributing to the leading-order term of the QCD pressure, 
$p_0$ of \eq\nr{eq:pHard}. Wiggly, dotted and full lines represent gluons, ghosts and quarks, respectively. 
Right panel: Two-loop Feynman diagrams contributing to $p_2$ of \eq\nr{eq:pHard}.}
\end{figure} }

\newcommand{\figureB}{%
\newcommand{\sbv}{\scalebox{0.72}}
\begin{figure}[tbp]
\begin{eqnarray*}
&&
 +{1\over16} \sbv{ \ToprVTT(\Agl,\Agl,\Agl,\Agl) }
 +{1\over48} \sbv{ \ToprVB(\Agl,\Agl,\Agl,\Agl) }
 +{1\over8} \sbv{ \ToprVBT(\Agl,\Agl,\Agl,\Agl,\Lgl) }
 -{1\over4} \sbv{ \ToprVBT(\Ahg,\Agl,\Agl,\Agl,\Lhg) }
 +{1\over8} \sbv{ \ToprVV(\Agl,\Agl,\Lgl,\Lgl,\Lgl) }
 +{1\over24} \sbv{ \ToprVM(\Agl,\Agl,\Agl,\Lgl,\Lgl,\Lgl) }
\\[3mm]&&
 -{1\over3} \sbv{ \ToprVM(\Agl,\Ahg,\Agl,\Lhg,\Lhg,\Lgl) }
 -{1\over4} \sbv{ \ToprVM(\Agl,\Ahg,\Ahg,\Lgl,\Lhg,\Lhg) }
 +{1\over16} \sbv{ \ToprVBB(\Agl,\Lgl,\Agl,\Lgl,\Lgl,\Lgl) }
 -{1\over4} \sbv{ \ToprVBB(\Agl,\Lgl,\Ahg,\Lgl,\Lgl,\Lhg) }
 +{1\over4} \sbv{ \ToprVBB(\Ahg,\Lgl,\Ahg,\Lgl,\Lhg,\Lhg) }
 -{1\over2} \sbv{ \ToprVBB(\Agl,\Lhg,\Agl,\Lhg,\Lhg,\Lhg) }
\\[3mm]&&
 -{1\over4} \sbv{ \ToprVBT(\Auq,\Agl,\Agl,\Agl,\Luq) }
 -{1\over4} \sbv{ \ToprVM(\Agl,\Auq,\Auq,\Lgl,\Luq,\Luq) }
 -{1\over3} \sbv{ \ToprVM(\Agl,\Auq,\Agl,\Luq,\Luq,\Lgl) }
 -{1\over4} \sbv{ \ToprVBB(\Agl,\Lgl,\Auq,\Lgl,\Lgl,\Luq) }
 +{1\over2} \sbv{ \ToprVBB(\Auq,\Lgl,\Ahg,\Lgl,\Luq,\Lhg) }
 -{1\over2} \sbv{ \ToprVBB(\Agl,\Luq,\Agl,\Luq,\Luq,\Luq) }
 +{1\over4} \sbv{ \ToprVBB(\Auq,\Lgl,\Auq,\Lgl,\Luq,\Luq) }
\end{eqnarray*}
\vspace*{-6mm} 
\caption{\la{fig:3loop} Three-loop Feynman diagrams contributing to $p_3$ of \eq\nr{eq:pHard}. 
The first two lines depict purely bosonic diagrams making up $p_3(\Nf=0)$, 
while the third line shows the diagrams proportional to $\Nf$ and $\Nf^2$ (last diagram only).
Notation as in \fig\ref{fig:12loop}.}
\end{figure} }

\newcommand{\figureC}{%
\newcommand{\sbk}{\scalebox{0.6}}
\begin{figure}[tbp]
\centering 
\begin{eqnarray*}
&&\!\!\!\!\!{}
 +{1\over24} \sbk{\TopfVBT(\Agl,\Agl,\Agl,\Agl,\Agl,\Agl) }
 +{1\over32} \sbk{\TopfVTTTT(\Agl,\Agl,\Agl,\Agl,\Agl,\Agl) }
 +{1\over48} \sbk{\TopfVTTT(\Agl,\Agl,\Agl,\Agl,\Agl,\Agl) }\!\!
 +{1\over48} \sbk{\TopfVT(\Agl,\Agl,\Agl,\Lgl,\Lgl,\Lgl) }
 +{1\over16} \sbk{\TopfVVTa(\Agl,\Agl,\Agl,\Agl,\Lgl,\Lgl,\Agl) }
 +{1\over4} \sbk{\TopfVVTb(\Agl,\Lgl,\Agl,\Agl,\Lgl,\Lgl,\Agl) }
 +{1\over16} \sbk{\TopfVSTT(\Agl,\Lgl,\Agl,\Agl,\Agl,\Agl,\Agl) }
\\[2mm]&&\!\!\!\!\!{}
 -{1\over8} \sbk{\TopfVSTT(\Ahg,\Lgh,\Agl,\Agl,\Agl,\Agl,\Agl) }
 +{1\over16} \sbk{\TopfVTST(\Agl,\Agl,\Agl,\Lgl,\Agl,\Agl,\Agl) }
 +{1\over16} \sbk{\TopfVBTT(\Agl,\Agl,\Agl,\Agl,\Agl,\Agl,\Agl) }
 -{1\over8} \sbk{\TopfVBTT(\Agl,\Agl,\Agl,\Agl,\Ahg,\Ahg,\Agl) }
 +{1\over8} \sbk{\TopfVN(\Agl,\Agl,\Lgl,\Lgl,\Lgl,\Lgl,\Lgl) }
 +{1\over8} \sbk{\TopfVB(\Agl,\Agl,\Agl,\Agl,\Agl,\Agl,\Lgl) }
 +{1\over16} \sbk{\TopfVU(\Agl,\Agl,\Agl,\Agl,\Lgl,\Lgl,\Lgl) }
\\[2mm]&&\!\!\!\!\!{}
 +{1\over24} \sbk{\TopfVBBa(\Agl,\Agl,\Agl,\Agl,\Agl,\Agl,\Agl) }
 -{1\over12} \sbk{\TopfVBBa(\Agl,\Agl,\Agl,\Agl,\Agl,\Ahg,\Agh) }
 +{1\over8} \sbk{\TopfVMT(\Agl,\Agl,\Agl,\Agl,\Lgl,\Lgl,\Lgl,\Agl) }
 -{1\over4} \sbk{\TopfVMT(\Agl,\Ahg,\Ahg,\Agl,\Lgh,\Lgl,\Lgh,\Agl) }
 -{1\over2} \sbk{\TopfVMT(\Agl,\Ahg,\Agl,\Agl,\Lgh,\Lhg,\Lgl,\Agl) }
 +{1\over16} \sbk{\TopfVBBT(\Agl,\Agl,\Agl,\Agl,\Agl,\Agl,\Agl,\Agl) }
 -{1\over4} \sbk{\TopfVBBT(\Agl,\Agl,\Agl,\Agl,\Agl,\Ahg,\Ahg,\Agl) }
\\[2mm]&&\!\!\!\!\!{}
 +{1\over4} \sbk{\TopfVBBT(\Agl,\Agl,\Agl,\Ahg,\Ahg,\Ahg,\Ahg,\Agl) }
 +{1\over8} \sbk{\TopfVBST(\Agl,\Lgl,\Lgl,\Lgl,\Lgl,\Agl,\Agl,\Agl) }
 -{1\over4} \sbk{\TopfVBST(\Agh,\Lhg,\Lgl,\Lgl,\Lgl,\Agl,\Agl,\Agl) }
 -{1\over2} \sbk{\TopfVBST(\Agl,\Lhg,\Lgh,\Lgh,\Lhg,\Agl,\Agl,\Agl) }
 +{1\over4} \sbk{\TopfVVBb(\Lgl,\Lgl,\Agl,\Lgl,\Lgl,\Lgl,\Lgl,\Agl) }
 -{1\over2} \sbk{\TopfVVBb(\Lgl,\Lgl,\Agl,\Lgl,\Lgl,\Lgl,\Lhg,\Ahg) }
 +{1\over4} \sbk{\TopfVV(\Agl,\Agl,\Agl,\Lgl,\Lgl,\Lgl,\Lgl,\Lgl) }
\\[2mm]&&\!\!\!\!\!{}
 -{1\over2} \sbk{\TopfVV(\Agl,\Agl,\Ahg,\Lgl,\Lhg,\Lhg,\Lgl,\Lgl) }
 +{1\over16} \sbk{\TopfVVBa(\Lgl,\Agl,\Agl,\Lgl,\Agl,\Agl,\Lgl,\Agl) }
 -{1\over8} \sbk{\TopfVVBa(\Lgl,\Agl,\Agl,\Lgl,\Agl,\Agl,\Lhg,\Ahg) }
 +{1\over8} \sbk{\TopfVW(\Agl,\Agl,\Agl,\Agl,\Lgl,\Lgl,\Lgl,\Lgl) }
 -{1\over4} \sbk{\TopfVW(\Ahg,\Ahg,\Ahg,\Ahg,\Lgl,\Lgl,\Lgl,\Lgl) }
 +{1\over32} \sbk{\TopfVSS(\Agl,\Lgl,\Agl,\Agl,\Agl,\Agl,\Lgl,\Agl) }
 -{1\over8} \sbk{\TopfVSS(\Agl,\Lgl,\Agl,\Agl,\Agl,\Agl,\Lhg,\Ahg) }
\\[2mm]&&\!\!\!\!\!{}
 +{1\over8} \sbk{\TopfVSS(\Agh,\Lhg,\Agl,\Agl,\Agl,\Agl,\Lhg,\Ahg) }
 +{1\over72} \sbk{\TopfVX(\Agl,\Agl,\Lgl,\Lgl,\Lgl,\Lgl,\Lgl,\Lgl,\Lgl) }
 -{1\over4} \sbk{\TopfVX(\Agl,\Agl,\Lgh,\Lgl,\Lgl,\Lhg,\Lhg,\Lgl,\Lhg) }
 -{1\over6} \sbk{\TopfVX(\Agl,\Agl,\Lgh,\Lhg,\Lhg,\Lhg,\Lhg,\Lhg,\Lgl) }
 +{1\over16} \sbk{\TopfVBSB(\Agl,\Lgl,\Lgl,\Lgl,\Agl,\Lgl,\Lgl,\Lgl,\Lgl) }
 -{1\over4} \sbk{\TopfVBSB(\Agl,\Lgl,\Lgl,\Lhg,\Ahg,\Lgl,\Lgl,\Lgl,\Lgl) }
 -{1\over2} \sbk{\TopfVBSB(\Agl,\Lgh,\Lhg,\Lgl,\Agl,\Lhg,\Lgh,\Lgl,\Lgl) }
\\[2mm]&&\!\!\!\!\!{}
 -{1\over2} \sbk{\TopfVBSB(\Agl,\Lhg,\Lgl,\Lgl,\Ahg,\Lgh,\Lhg,\Lgh,\Lhg) }
 +{1\over4} \sbk{\TopfVBSB(\Agh,\Lhg,\Lgl,\Lhg,\Ahg,\Lgl,\Lgl,\Lgl,\Lgl) }
 +{1} \sbk{\TopfVBSB(\Agl,\Lgh,\Lhg,\Lhg,\Ahg,\Lhg,\Lgh,\Lgl,\Lgl) }
 +{1\over12} \sbk{\TopfVH(\Agl,\Agl,\Lgl,\Lgl,\Lgl,\Lgl,\Lgl,\Lgl,\Lgl) }
 -{1\over2} \sbk{\TopfVH(\Agl,\Agl,\Lgh,\Lgl,\Lgl,\Lhg,\Lhg,\Lgl,\Lhg) }
 -{1\over3} \sbk{\TopfVH(\Agl,\Ahg,\Lgl,\Lgl,\Lgl,\Lgl,\Lhg,\Lhg,\Lgl) }
 -{1} \sbk{\TopfVH(\Agl,\Ahg,\Lhg,\Lhg,\Lgh,\Lgh,\Lgl,\Lgl,\Lgl) }
\\[2mm]&&\!\!\!\!\!{}
 -{1\over2} \sbk{\TopfVH(\Agl,\Agl,\Lgh,\Lgh,\Lhg,\Lhg,\Lhg,\Lhg,\Lgl) }
 +{1\over6} \sbk{\TopfVH(\Ahg,\Ahg,\Lgl,\Lgl,\Lhg,\Lhg,\Lhg,\Lhg,\Lgl) }
 +{1\over6} \sbk{\TopfVH(\Ahg,\Agh,\Lgl,\Lgl,\Lhg,\Lhg,\Lgh,\Lgh,\Lgl)  } 
 +{1\over8} \sbk{\TopfVMB(\Lgl,\Agl,\Agl,\Lgl,\Lgl,\Lgl,\Lgl,\Lgl,\Agl) }
 -{1\over4} \sbk{\TopfVMB(\Lgl,\Agl,\Agl,\Lgl,\Lgl,\Lgl,\Lgl,\Lhg,\Ahg) }
 -{1} \sbk{\TopfVMB(\Lgh,\Agl,\Agl,\Lhg,\Lhg,\Lgl,\Lhg,\Lgl,\Ahg) }
 -{1\over2} \sbk{\TopfVMB(\Lgl,\Ahg,\Agl,\Lgl,\Lgh,\Lhg,\Lgl,\Lgl,\Agl) }
 -{1} \sbk{\TopfVMB(\Lgh,\Agl,\Ahg,\Lhg,\Lhg,\Lgh,\Lgl,\Lgl,\Ahg) }
\\[2mm]&&\!\!\!\!\!{}
 -{1\over4} \sbk{\TopfVMB(\Lgl,\Ahg,\Ahg,\Lgl,\Lgh,\Lgl,\Lgh,\Lgl,\Agl) }
 +{1} \sbk{\TopfVMB(\Lgl,\Ahg,\Agl,\Lgl,\Lgh,\Lhg,\Lgl,\Lhg,\Ahg) }
 +{1\over2} \sbk{\TopfVMB(\Lgl,\Ahg,\Ahg,\Lgl,\Lgh,\Lgl,\Lgh,\Lhg,\Ahg) }
 +{1\over48} \sbk{\TopfVBBb(\Agl,\Agl,\Agl,\Agl,\Agl,\Agl,\Agl,\Agl,\Agl) }
 -{1\over8} \sbk{\TopfVBBb(\Agl,\Agl,\Agl,\Agl,\Agl,\Ahg,\Ahg,\Agl,\Agl) }
 -{1\over3} \sbk{\TopfVBBb(\Ahg,\Ahg,\Ahg,\Ahg,\Agl,\Ahg,\Agl,\Ahg,\Agl) }
 +{1\over4} \sbk{\TopfVBBb(\Agl,\Agl,\Agl,\Ahg,\Ahg,\Agl,\Agl,\Ahg,\Ahg) }
 -{1\over6} \sbk{\TopfVBBb(\Agl,\Agl,\Agl,\Ahg,\Ahg,\Ahg,\Ahg,\Ahg,\Ahg) }
\end{eqnarray*}
\vspace*{-6mm} 
\caption{\la{fig:4loop} Feynman diagrams contributing to the 4-loop pressure of pure gauge theory, $p_4(\Nf=0)$.}
\end{figure} }

\newcommand{\figureD}{%
\begin{figure}[tbp]
\centering 
\begin{align*}
\TLfig{511}
\TLfig{1022}
\TLfig{1011}
\TLfig{1020}
\TLfig{1009}
\TLfig{1010}
\TLfig{1012}
\TLfig{1016}
\\
\TLfig{952}
\TLfig{978}
\TLfig{993}
\TLfig{1008}
\TLfig{841}
\TLfig{961}
\TLfig{992}
\TLfig{960}
\end{align*}
\vspace*{-4mm} 
\caption{\la{fig:sectors}Set of 16 unique sector representatives for the 4-loop (sum-) integrals corresponding 
to the momentum family \eq\nr{eq:family}. 
The enumeration corresponds to the decimal representation of the respective binary footprint of propagators.}
\end{figure} }

\title{\boldmath The $g^6$ pressure of hot Yang-Mills theory: Canonical form of the integrand}

\author[a,b]{Pablo~Navarrete}
\author[a]{and York~Schr\"oder}

\affiliation[a]{Centro de Ciencias Exactas, Departamento de Ciencias B\'asicas, Universidad del B\'io-B\'io, 
Avenida Andr\'es Bello 720, Chill\'an, Chile}
\affiliation[b]{Department of Physics and Helsinki Institute of Physics, P.O.\ Box 64, University of Helsinki, Finland}

\emailAdd{pablo.navarrete@helsinki.fi}
\emailAdd{yschroder@ubiobio.cl}

\arxivnumber{2408.15830}

\abstract{We present major progress towards the determination of the last missing piece for the pressure 
of a Yang-Mills plasma at high temperatures at order $g^6$ in the strong coupling constant.
This order is of key importance due to its role in resolving the long-standing infrared problem of finite-temperature
field theory within a dimensionally reduced effective field theory setup.
By systematically applying linear transformations of integration variables, or momentum shifts, 
we resolve equivalences between different representations of Feynman sum-integrals
on the integrand level, transforming those into a canonical form.
At the order $g^6$, this results in reducing a sum of ${\cal O}(100000)$ distinct
sum-integrals which are produced from all four-loop vacuum diagrams down to merely 21.
Furthermore, we succeed to map 11 of those onto known lower-loop structures.
This leaves only 10 genuine 4-loop sum-integrals to be evaluated, 
thereby bringing the finalization of three decades of theoretical efforts within reach.}

\begin{document}
\maketitle
\flushbottom

%
\section{Introduction}
\la{se:intro}

It had been argued \cite{Linde:1978px,Linde:1980ts,Gross:1980br}
and indeed widely accepted \cite{Landsman:1989be,Blaizot:2001nr},
that a weak-coupling expansion of hot QCD can not be consistently performed.
The argument rests on the observation that transverse (``chromomagnetic'') gluonic modes
that remain unscreened when propagating through the hot plasma 
cause an infrared problem, which for the pressure arises at four loops and beyond.

Based on more recent understanding of effective field theory, however, 
a roadmap to evade this infrared problem in the setting of dimensionally reduced effective 
theories \cite{Ginsparg:1980ef,Appelquist:1981vg,Kajantie:1995dw}
has been laid out in a seminal paper \cite{Braaten:1995jr}.
This has spurred enormous theoretical efforts \cite{Kajantie:2000iz,Kajantie:2001hv,Kajantie:2003ax,Hietanen:2004ew,
Giovannangeli:2005rz,Schroder:2005zd,Laine:2006cp,DiRenzo:2006nh,DiRenzo:2008en,Ghisoiu:2015uza}, 
completing most but one of the required steps.
The one missing piece of information is a well-defined perturbative coefficient in hot QCD, 
at the four-loop level.

As the root cause of the infrared problem lies in the gluonic sector of QCD, 
we set the number of fermions $\Nf=0$ in this work and focus on the pure Yang-Mills case, 
as already suggested by the manuscript's title. 
The addition of fermions does not present any new fundamental problems, 
and might be classified as more of a bookkeeping exercise.
Indeed, weak-coupling results for full QCD up to $g^5$ have been known for a long time 
\cite{Shuryak:1977ut,Chin:1978gj,Kapusta:1979fh,Toimela:1982hv,Arnold:1994ps,Arnold:1994eb,Zhai:1995ac},
and been partially extended to the four-loop level at large $\Nf$ \cite{Gynther:2009qf},
with those $\Nf\neq0$ corrections generally exhibiting very good convergence properties,
in contrast to the situation at $\Nf=0$ \cite{Zhai:1995ac,Braaten:1995jr,Kajantie:1997tt,Kajantie:2002wa,Blaizot:2003iq}.

There are order-of-magnitude numerical estimates at $\Nf=0$
of the above-mentioned missing perturbative coefficient \cite{Kajantie:2002wa,Laine:2006cp,Schroder:2006vz} 
based on comparisons with non-perturbative lattice Monte Carlo simulations 
of pure Yang-Mills theory \cite{Boyd:1996bx,Beinlich:1997ia,CP-PACS:1999eop,Borsanyi:2012ve}
at a scale where both approaches might still be applicable. 
To appreciate why a diagrammatic evaluation of this coefficient is still missing to date,
it is instructive to simply count the number of terms that appear in the (sum-) integrand
of purely gluonic 4-loop vacuum Feynman diagrams, after multiplying out 
the vertex- (6 terms per gluon 3-vertex) and propagator- (2 terms per gluon propagator) structures.
For example, diagrams that contain gluon-3-vertices only
(see lines 6 and 7 of \fig\ref{fig:4loop} for concrete examples)
contain 6 vertices and 9 propagators, leading to $6^6\times2^9\approx $ 24 million terms each
before combining equal terms. 

The main focus of the present paper is to tackle this open perturbative problem head on,
organizing the enormous number of terms that appear in the 4-loop expansion of the Yang-Mills pressure
in general covariant gauges.
To this end, we propose an algorithm that makes systematic use of momentum shift invariance, 
enabling us to define a canonical form on the integrand level.
As a result, we manage to eliminate redundancy among equivalent (sum-) integral representations,
observing an enormous reduction in expression size of the 4-loop problem, 
and even obtaining gauge parameter independence as a welcome by-product.
Furthermore, we map some of the remaining canonicalized integrals onto known factorized cases, 
and close by precisely defining a few open calculations that would be needed to finish the roadmap that
had been sketched almost three decades ago \cite{Braaten:1995jr}.

The structure of the remainder of the paper is as follows. 
We start by defining the observable in question and introduce some useful related notation in \se\ref{se:setup}.
Details of our computation regarding the canonicalization and the mapping onto factorized sum-integrals
are explained in \secs\ref{se:reduction} and \ref{se:evals}, respectively.
We conclude in \se\ref{se:conclu}, and devote two appendices to collect information on lower-loop coefficients
as well as relevant known cases of sum-integral evaluations.

%
\section{Setup}
\label{se:setup}

The above-mentioned dimensionally reduced effective theory setup 
that enables a high-temperature weak-coupling expansion of the full QCD 
pressure\footnote{The pressure is an equilibrium quantity, 
so we work in the imaginary-time formalism \cite{Landsman:1986uw} 
with Euclidean Lagrange density $L_\rmi{QCD}^\rmi{E}$ that results
after rotating $t\rightarrow i\tau$. Bosons (fermions) are (anti)periodic in $\tau$.}
\ba\la{eq:pDef}
p_\rmi{QCD}(T) &=& \lim_{V\rightarrow\infty} \frac{T}{V}\, \ln \int\!{\cal D}[A,\bar\psi,\psi]\,
\exp\Big(-\int_0^{1/T}\!\!\!\!{\rm d}\tau\int\!{\rm d}^d{\vec x}\;L_\rmi{QCD}^\rmi{E}\Big)
\ea
in terms of distinct contributions from three formally well-separated (so-called hard, soft and ultrasoft) momentum scales
has been described in many places 
\cite{Braaten:1995jr,Kajantie:2002wa,Schroder:2006vz}.
It leads to the three-term separation $p_\rmi{QCD}=p_\rmi{hard}+p_\rmi{soft}+p_\rmi{ultrasoft}$,
where the first two are perturbative and the last is not.
We will not repeat the details of this systematic decomposition here, 
but focus directly on the key remaining open problem,
being the determination of the 4-loop term in a strict\footnote{That is, disregarding the known 
infrared divergences that persist even in the sum of all bare diagrams at that order, but regularizing them
together with the ultraviolet ones within dimensional regularization.} 
perturbative expansion of $p_\rmi{QCD}$ in terms of the bare gauge coupling $\gb^2$.
This specific coefficient arises from the hard momentum scales 
and is sometimes referred to as $\beta_\rmi{E1}$ in the literature \cite{Kajantie:2002wa}.

\figureA

For an SU($\Nc$) gauge group, we write the relevant perturbative expansion as
\ba\la{eq:pHard}
p_\rmi{hard}^\rmi{bare} &=& 
p_0 + \gb^2\Nc\,p_2 + \gb^4\Nc^2\,p_3 +\gb^6\Nc^3\,p_4 +{\cal O}(\mbox{5-loop}) \;. 
\ea
The free gauge theory gives rise to the ideal-gas term 
$p_0=\frac{\pi^2\,T^4}{45} \big[ (\Nc^2-1) +\frac74\,\Nc\Nf \big]$ 
whose value can be obtained\footnote{Famously, the ghost  
corrects the overcounting in the partition function
due to the gluonic degrees of freedom as $p_0=\frac{\pi^2\,T^4}{45} \,(\Nc^2-1) \big[2-1+{\cal O}(\Nf)\big]$,
where the contributions shown in square brackets originate from the first three diagrams of \fig\ref{fig:12loop}, in order.} 
from the first three 1-loop diagrams shown in \fig\ref{fig:12loop}.
Interactions then give rise to the 2-loop correction $p_2$ \cite{Shuryak:1977ut,Chin:1978gj},
the 3-loop term $p_3$ \cite{Arnold:1994ps,Arnold:1994eb}, 
and $p_4$ at four loops which contains $\beta_\rmi{E1}$. 
We show all Feynman diagrams that make up $p_2$ and $p_3$ in \figs\ref{fig:12loop} and \ref{fig:3loop},
while for $p_4$ we display only the pure-gauge part in \fig\ref{fig:4loop}.
A concise review of the known coefficients up to three loops is presented in \cite{Nishimura:2012ee}.
Below, we provide analytic expressions for the $p_n$, 
for $d$ dimensions and at $\Nf=0$, where $p_4(\Nf=0)$ constitutes the main new result of our paper.
For completeness, we note that in full QCD the 4-loop term receives four separate contributions that may be indexed
by the respective power of $\Nf$ as $p_4=p_{40}+\Nf\,p_{41}+\Nf^2\,p_{42}+\Nf^3\,p_{43}$. 
Of these, only $p_{43}$ can be considered known since it can be extracted at large $\Nf$ \cite{Moore:2002md,Ipp:2003zr}
or from a single Feynman diagram \cite{Gynther:2009qf}.
To re-iterate, we focus on $p_{40}$ here.

\figureB

Let us fix some notation. 
Due to the compact temporal integration domain in \eq\nr{eq:pDef}, 
Fourier transforming to momentum space leads to a tower of discrete (Matsubara) frequencies that 
need to be summed over. 
We employ Euclidean four-momenta $K=(K_0,{\vec k})$, 
where the components $K_0 = 2n_k\pi T$ are bosonic Matsubara frequencies with $n_k\in\setZ$. 
Massless (bosonic) propagators are then $1/[K^2] = 1/[(K_0)^2 + {\vec k}^2]$, and we define the sum-integral symbol as
\ba
\Tint{K} \equiv T\sum_{n_k\in\setZ} \int\frac{{\rm d}^d{\vec k}}{(2\pi)^d} \;,
\ea
where $d=3-2\ep$ reflects dimensional regularization. 

To facilitate the analysis of all momentum-space vacuum Feynman (sum-) integrals that contribute to $p_4$,
we use a fixed set of massless 4-loop propagators 
constructed from squares of the ten linear combinations $P_a^\mu$ of the four loop momenta 
$K_1^\mu\dots K_4^\mu$ given by (we omit Lorentz indices and write $K_{i-j}\equiv K_i-K_j$)
\ba \la{eq:family}
\{P_1,\dots,P_{10}\} = \{K_1,K_2,K_3,K_4,K_{1-4},K_{2-4},K_{3-4},K_{1-2},K_{1-3},K_{1-2-3}\} \;.
\ea
This allows to express any scalar bosonic 4-loop vacuum sum-integral, 
as needed for the diagrams shown in \fig\ref{fig:4loop}, as a list of 10 numbers $s_a\in\setZ$ as
\ba\la{eq:listNotation}
\ci(s_1,\dots,s_{10}) &\equiv& \Tint{K_1} \Tint{K_2} \Tint{K_3} \Tint{K_4} I(s_1,\dots,s_{10}) \;,\\
I(s_1,\dots,s_{10}) &\equiv& \frac1{(P_1\ccdot P_1)^{s_1}\dots (P_{10}\ccdot P_{10})^{s_{10}}} 
\la{eq:integrands} \;,
\ea
where positive (negative) indices $s_a$ correspond to propagators (numerators).
Our notation distinguishes sum-integrals $\ci$ from their integrands $I$.
For 4-loop contributions that factorize into lower-loop factors, it will be useful to define
the 3-loop momentum family
\ba \la{eq:family3loop}
\{P_1,\dots,P_{6}\}=\{K_1,K_2,K_3,K_{1-2},K_{1-3},K_{2-3}\}
\ea
and represent 3-loop scalar sum-integrals and their integrands as 6-element lists,
in complete structural analogy to the 4-loop case.

\figureC

%
\section{Reduction to canonical form}
\la{se:reduction}

\newcommand{\code}[1]{{\tt{#1}}}

The core of our calculation uses automatized tools as far as possible, and proceeds as follows. 
All Feynman diagrams are generated by \code{Qgraf} \cite{Nogueira:1991ex,Nogueira:2006pq}
and mapped onto our preferred momentum conventions of \eq\nr{eq:family}.
We then use the computer algebra package \code{FORM} \cite{Vermaseren:2000nd,Ueda:2020wqk}
to plug in Feynman rules (we work in general covariant gauges with gauge parameter $\xi$), 
perform the color and Lorentz algebra, and express all resulting scalar
sum-integrals in the form of lists of integers as defined in \eq\nr{eq:listNotation}.
These lists are analyzed with \code{Mathematica} \cite{mathematica} code,
which exploits the sum-integrals' invariance under linear transformations of loop momentum variables 
and produces replacement rules that allow to bring the lists into a canonical form.
Finally, these replacement rules are applied to the sum-integrals at hand and the contributions from 
all diagrams are summed up, at which point we observe a very substantial reduction of terms.
We will comment on some of these steps in the following section.
A reader not interested in technical details might skip right ahead to \eq\nr{eq:p4can} where
we present the form of our result.

%
\subsection{Mapping}
\la{se:qgraf}

\newcommand{\secID}{{\rm id}}
\newcommand{\Upoly}{{\cal U}}

In practice, we start by generating all diagrams that contribute to $p_4$ by \code{Qgraf}.
As a minor detail, we generate two-point functions that result from cutting a line in a (connected, 
one-particle irreducible) vacuum diagram and glue the two external lines back together 
with a propagator, correcting for the symmetry factor as explained e.g.\ in \cite{Kajantie:2001hv}.
We perform this step in two different ways: by gluing together all (connected) gluonic\footnote{Since ghosts 
(and quarks) do not self-interact, a vacuum diagram at two loops or higher is guaranteed to have at least one
gluon propagator, such that we are guaranteed (even in full QCD) to obtain all possible vacuum diagrams from 
only considering gluon self-energies.}
self-energies, or by following the somewhat more refined procedure of \cite{Kajantie:2001hv} that consists
in generating 2-particle irreducible skeleton vacuum graphs separately from 2-particle reducible ring diagrams
(the relevant self-energy selection options within \code{Qgraf} are \code{onepi}, \code{nosigma} and \code{sigma}).
Both ways give rise to the set of diagrams depicted in \fig\ref{fig:4loop}, with precisely the same combinatorical factors,
but with different internal momentum routings. 
We keep both sets in our computation, to provide a strong check on internal consistency and completeness 
of our reduction algorithm in the end of our calculation.

Suppose we are now given (by a collaborator, or a program such as \code{Qgraf}) a 4-loop vacuum Feynman diagram 
that employs a momentum list $\bar P$, not necessarily coinciding with our choice $P$ of \eq\nr{eq:family}.
We then need replacement rules that map the $\bar P$ onto (a subset of) the $P$, in order to be able
to express all (sum-) integrals that arise from the given diagram in our list notation of \eq\nr{eq:listNotation}.
Such replacement rules can be generated in an automatized way by comparing the
so-called graph (or first Symanzik) polynomials $\Upoly$ of given momentum lists $\bar P$ and $P$. 
For an $L$\/-loop graph, $\Upoly$ is a homogeneous polynomial of degree $L$ in the variables $x_a$ 
that are attached to the edges. It encodes the underlying graph
in a unique way, up to re-numberings of the $x_a$.
It plays an important role in parametric representations of Feynman integrals 
(see e.g.\ the recent textbook \cite{Weinzierl:2022eaz}), 
but we can as well make use of $\Upoly$ to detect graph isomorphisms, aided
by a canonical ordering for such polynomials, based on ideas outlined in \cite{Pak:2011xt,Hoff:2015kub}.
Once we have detected that two graph polynomials are equivalent, we can reverse the canonical ordering
to match pairs of momenta from the sets $\bar P$ and $P$, and obtain the replacement rules.
So we can guarantee that any input can be mapped onto our choice of momentum family \eq\nr{eq:family}.

It is now easy to plug in the Feynman rules, perform the color and Lorentz traces (we use \code{FORM} for these steps) 
and arrive at scalar vacuum sum-integrals represented in list notation \eq\nr{eq:listNotation}. 
Note that we are dealing with vacuum diagrams; there are no external momenta and all (Lorentz and group) 
indices contract, guaranteeing a scalar (sum-) integral as outcome. Furthermore, owing to the fact that our
family is complete in the sense that all scalar products $K_j^\mu K_k^\mu$ can be uniquely represented 
as (inverse) propagators, all structures can be represented as integrals $\ci$ of \eq\nr{eq:listNotation}.

Recalling \eq\nr{eq:pHard}, the hard contribution $p_4$ to the bare pressure of pure SU($\Nc$) Yang-Mills theory 
at order $O(\gb^6)$ follows from the 65 vacuum Feynman diagrams containing gluons and ghosts 
as displayed in \fig\ref{fig:4loop}. 
After applying the momentum mapping described above and in general covariant gauge, it takes the form
\ba \la{eq:p4full}
p_4(d,T) &=& (\Nc^2-1) \sum_{n=1}^{N} \tilde{c}_n(d,\xi)\;\tilde\ci_n(d,T),
\ea
where the $\tilde{c}_n$ are polynomial coefficient functions in the dimension $d$ and the gauge parameter $\xi$
(where the latter occurs only up to $\xi^6$), and $N\sim {\cal O}(100000)$ is the number 
of distinct\footnote{The exact value of $N$
depends somewhat on how the Feynman diagrams were constructed and labelled. For example, our two different
methods of generating the diagrams as explained above result in $N=175998$ and $N=96928$,
respectively. Running both different sets through the same canonicalization algorithm and arriving at identical
canonical forms can then be interpreted as a very strong check on the calculation.} scalarized sum-integrals $\tilde\ci_n(d,T)$ 
in the form of \eq\nr{eq:listNotation}, as obtained after mapping all sum-integrals produced by the complete 
set of diagrams onto our conventions.

%
\subsection{Canonical form}
\la{se:canonical}

There is enormous freedom in writing down a sum-integral.
The value of $\ci$ does not depend on the particular propagator momentum choice.
It is invariant under reparameterizations of the loop momenta.
In particular, we can allow for linear changes of variables $K\rightarrow\bar K=SK$
where $S$ are unimodular $L\times L$ matrices (i.e.\ square integer matrices with $\det S=\pm1$).
The $S$ are clearly invertible;
the transformations have unit Jacobian determinant and therefore leave the measure $\Tinti{}$ invariant;
hence such shifts $S$ induce equivalence relations
directly among the integrands $I$ of \eq\nr{eq:integrands}.

We can once more make use of the graph polynomials $\Upoly$ to detect graph isomorphisms.
Now, we are interested in the different ways that the same $\ci$ can be represented by different integrands $I$.
Systematically applying momentum shifts $S$ to subsets of the full family $P$ 
and then comparing graphs on the basis of their canonically ordered graph polynomials $\Upoly$ as
explained in \se\ref{se:qgraf}, we can define equivalence classes, and choose unique representatives.
In \fig\ref{fig:sectors}, we display the result of such a classification on the basis of propagator lines
(i.e.\ the positive list entries $s_a$ of an integrand $I(s_1,\dots,s_{10})$, defining a so-called sector of the full family).

\figureD

The core of our canonicalization approach can now be stated as follows: 
identify equivalence classes on the integrand level,
pick a unique representative for each class (e.g.\ the ``smallest’’ according to some ordering relation $\ci_1\prec \ci_2$),
and exploit the shifts $S$ to map any (linear combination of) $\ci$ onto those representatives,
the result of which we call the canonical form of an expression.

This concept is of course far from original, 
and it is indeed heavily used in perturbative quantum field theory at zero temperature,
in particular in the context of algorithmic integration-by-parts (IBP) approaches \cite{Chetyrkin:1981qh,Laporta:2000dsw}.
As far as we are aware, however, this machinery has not yet been systematically applied
to sum-integrals on a large scale.

As can be inferred from the above, part of our strategy relies heavily on matrix operations. 
For ease of implementation, we have therefore opted to use a commercial computer algebra system
(Wolfram Mathematica \cite{mathematica}) to develop key parts of our setup.
We regard this choice as a proof-of-principle only, 
and stress that the strategy can (and maybe should) also be implemented on freely available software.

On the technical side, the large linear systems we encounter can be dealt with in an efficient way
using sparse matrix representations (\code{SparseArray}), 
exact integer-matrix algorithms such as (\code{HermiteDecomposition}), 
and exploiting the block-diagonal form of the systems that is induced by the sector- and sub-sector 
(as obtained by deleting a line of the parent sector) structure of our sum-integrals.

%
\subsection{Bare pressure at $g^6$}
\la{se:result}

Let us now return to our physics problem at hand and simplify the 4-loop hard pressure of \eq\nr{eq:p4full}.
Following the strategy laid out in \se\ref{se:canonical}, the large redundancy present in the vector 
space spanned by the set $\{\tilde\ci_n(d,T)\}_{n=1}^N$ is alleviated upon solving a system of linear equations, 
generated by a systematic application of momentum shifts. 
In the process, we witness a significant reduction of the number of sum-integrals in need of evaluation, 
in addition to the explicit cancellation of all factors of $\xi$, serving as a non-trivial check of the correctness 
of the procedure. 
Equation~\nr{eq:p4full} then collapses to
\newcommand{\mi}{\!-\!}
\newcommand{\pl}{\!+\!}
\ba
p_4(d,T) &=& -\frac{(\Nc^2-1)}{72}  \sum_{n=1}^{21} c_n(d)\,\ci_n(d,T) \\
&=& -\frac{(\Nc^2-1)}{72}\,\Big\{
(1939-852d+63d^2+2d^3)\ci_6
+6(281-180d+39d^2+4d^3)\ci_9
\nonumber\\&+&
2(d\mi 1)^3 \Big[ 9(d\mi 9)\ci_1 \pl 6(d\mi 5)\ci_2 \pl 18\ci_3 \mi 18\ci_4 \pl 12\ci_{10} \mi 48\ci_{11} 
\pl 27\ci_{12} \pl 36\ci_{13} \mi \ci_{16} \Big]
\nonumber\\&-&
3(d\mi 1)^2  \Big[ 8(7d\mi 31)\ci_5 +(d\pl 15)\ci_{15} +4\ci_{18} -4\ci_{19} -6\ci_{20} +4\ci_{21} \Big]
\nonumber\\&+&
18(41-19d+2d^2) \Big[ \ci_7 -4\ci_8 +\ci_{17} \Big]
-18(119-49d+2d^2)\ci_{14}
\;\Big\} \;. \la{eq:p4can}
\ea
The resulting set of 21 sum-integrals $\ci_n(d,T)$ can be understood as a basis in the sense of corresponding 
to sector representatives (see \fig \ref{fig:sectors}) of the minimal set of integrals\footnote{Finding a genuine 
linearly independent basis of so-called master integrals is only feasible after taking into account integration-by-parts 
relations as well. However, this is well understood only at $T=0$ \cite{Smirnov:2010hn}.} 
that are independent with respect to the graphs' internal symmetries, a property we refer to as canonical form.
We enumerate and discuss the $\ci_n$ in the next section.

As a further check on the canonicalization procedure developed and applied for reducing the four-loop term $p_4$
to its compact form \eq\nr{eq:p4can}, we can apply the same recipe for the two- and three-loop pressure,
see \figs\ref{fig:12loop} and \ref{fig:3loop} for the corresponding diagrams.
Details and analytic $d$\/-dimensional expressions for $p_0$, $p_2$ and $p_3$ of \eq\nr{eq:pHard} 
are given in \app\ref{se:123loop}.
It is reassuring to observe that those expressions coincide with the known results \cite{Braaten:1995jr,Nishimura:2012ee}.

While \eq\nr{eq:p4can} together with the explicit $\ci_n$ listed in the following section 
is the main result of this paper, we now proceed with the characterization and 
evaluation of (a subset of) the $\ci_n$, particularly those which are factorizable into known lower-loop factors.

%
\section{Sum-integral evaluations}
\la{se:evals}

The 21 sum-integrals $\ci_n(d,T)$ entering \eq\nr{eq:p4can} can be classified according to
the structure of the underlying propagator graph sector.
Eleven of them fall into the subset of sectors that decompose into lower-loop factors,
recall \fig\ref{fig:sectors} where we have displayed and labelled all such sectors. 
Below, we will frequently attach to each sum-integral the corresponding sector number 
in form of a subscript in square brackets. This is for convenience only, and does not change the
basic list notation of \eq\nr{eq:listNotation}. 
In fact, we will also indicate the number of lines of each sector in that subscript,
and omit the function arguments $(d,T)$ for brevity.

The factorizable sum-integrals can be readily evaluated, as they are either fully known 
(the case of one and two loops), or can be mapped onto known 3-loop cases, 
thereby leaving only 10 genuine (non-factorizable) 4-loop sum-integrals in need of evaluation.

We will now turn to evaluate all 11 factorizable sum-integrals that we observe in our sum \eq\nr{eq:p4can},
and finally list the 10 non-factorizable cases, for which we discuss possible evaluation strategies.

%
\subsection{Factorized sectors: one loop}
\la{se:eval1loop}

We obtain two sum-integrals fully factorized as four 1-loop factors, reading 
\ba
\ci_1 &\equiv& \ci(2,2,1,1,0,0,0,0,0,0)_{[4,960]}\;, \\
\ci_2 &\equiv& \ci(3,1,1,1,0,0,0,0,0,0)_{[4,960]}\;.
\ea
They can be expressed in terms of the analytically known 1-loop tadpoles of \eq\nr{eq:I} as
\ba
\ci_1= (I^0_2)^2\;(I^0_1)^2 \;,\quad 
\ci_2 = I^0_3\;(I^0_1)^3\;.
\ea

%
\subsection{Factorized sectors: two loops}
\la{se:eval2loop}

Three sum-integrals factorize onto 1- and 2-loop factors; they read
\ba
\ci_4 &\equiv& \ci(1,2,1,1,1,0,0,0,0,0)_{[5,992]}\;, \\
\ci_5 &\equiv& \ci(2,1,1,1,1,0,0,0,0,0)_{[5,992]}\;, \\
\ci_7 &\equiv& \ci(1,1,1,1,0,1,0,0,1,0)_{[6,978]}\;.
\ea
Using \eqs\nr{eq:I} and \nr{eq:L}, they can be written as
\ba
\ci_4 = \sumintL^{0,0,0}_{1,1,1}\; I^0_1\; I^0_2 \;,\quad 
\ci_5 = \sumintL^{0,0,0}_{2,1,1} \; (I^0_1)^2 \;, \quad 
\ci_7 = (\sumintL^{0,0,0}_{1,1,1})^2\;.
\ea
Applying the generic reduction formula of \cite{Davydychev:2023jto} to the 2-loop factors $L$ 
then results in
\ba \la{eq:L111}
\ci_4 = 0\;, \quad 
\ci_5 = - \frac{(I^0_1)^2\;(I^0_2)^2}{(d-5)(d-2)} \;, \quad 
\ci_7 = 0\;.
\ea

%
\subsection{Factorizable sectors: three loops}
\la{se:eval3loop}

Six of our sum-integrals are built from one 1-loop and one 3-loop structure, 
and we have in fact already discussed these cases earlier \cite{Navarrete:2022adz}. They are 
\ba
\ci_3 &\equiv& \ci(2,1,1,1,0,0,0,0,0,1)_{[5,961]} \;,\\
\ci_9 &\equiv& \ci(1,1,1,1,1,1,0,0,0,0)_{[6,1008]}\;, \\
\ci_{10} &\equiv& \ci(1,1,1,3,1,1,0,0,-2,0)_{[6,1008]}\;, \la{eq:I10} \\
\ci_{11} &\equiv& \ci(1,1,1,3,1,1,0,0,-1,-1)_{[6,1008]}\;, \\
\ci_{12} &\equiv& \ci(1,1,1,3,1,1,0,0,0,-2)_{[6,1008]}\;, \\
\ci_{13} &\equiv& \ci(2,1,1,2,1,1,0,-1,0,-1)_{[6,1008]}\;. \la{eq:I13}
\ea
Note that sector [1012] does not contribute.
The sum-integrals $\ci_3$ and $\ci_9$ do not have numerators and trivially decouple into two factors.  
In terms of the known 3-loop cases given in \eqs\nr{eq:J11} and \nr{eq:J1}, they evaluate to
\ba
\ci_3 = J_1\;I^0_1\;, \quad 
\ci_9 = J_{11}\;I^0_1 \;.
\ea
The four remaining sum-integrals of \eqs\nr{eq:I10}--\nr{eq:I13} contain numerator factors, 
necessitating tensor decomposition in order to decouple their 1- and 3-loop parts. 
Thanks to the presence of the 1-loop tadpole factor this is straightforward, and we can apply the general method
developed in \cite{Navarrete:2022adz}.
After furthermore employing simple 3-loop IBP reductions as well as the 2-loop 
reduction formula of \cite{Davydychev:2023jto}, this leads to \cite{Navarrete:2022adz}
\ba
\ci_{10} &=& \frac{16}{d(d-2)(d-7)}\,I^2_1\; I^0_3\; I^0_2\; I^0_1 
- \frac{(I^0_2)^2\;(I^0_1)^2}{(d-2)(d-5)} + \frac{4(d+1)}{d}\,V_4\; I^2_1\;, \la{eq:I10_2} \\
\ci_{11} &=& I^0_3\; (I^0_1)^3 + \frac{16}{d(d-2)(d-7)}\,I^2_1\; I^0_3\; I^0_2\; I^0_1 
+ \frac{(d+1)}{d(d-2)(d-5)^2} \,I^2_1\; (I^0_2)^3 \nonumber \\
&+& \frac{(d+1)}{d(d-5)}\, V_1\; I^2_1 - \frac{(4-7d+d^2)}{2d(d-5)}\, V_2\; I^2_1 
+ \frac{4(d+1)}{d}\, V_4\; I^2_1\;, \la{eq:I11_2} \\
\ci_{12} &=& J_{13}\; I^0_1 \la{eq:I12_2} 
+ \frac{32}{d(d-2)(d-7)}\,I^2_1\; I^0_3\; I^0_2\; I^0_1 
+ \frac{2(d+1)}{d(d-2)(d-5)^2}\,I^2_1\; (I^0_2)^3 \nonumber \\
&+&  \frac{2(d+1)}{d(d-5)}\,V_1\; I^2_1 
- \frac{(4-7d+d^2)}{d(d-5)}\,V_2\; I^2_1 
+ \frac{8(d+1)}{d}\,V_4\; I^2_1 \;,\\
\ci_{13} &=& \frac{1}{2}\,I^0_1 \biggl[ \frac{2(2d-11)}{(d-5)}\,(I^0_2)^2\; I^0_1 - J_1 - 2\,J_{11} - 4(d-4)\,J_{12} \biggl]\;. \la{eq:I13_2}
\ea
All the required 3-loop sum-integrals are known up to $O(\ep^0)$, and listed in \app \ref{se:3loop}.

%
\subsection{Non-factorized sectors}

The ten remaining sum-integrals fall in the category of genuine non-factorized sectors, and none of them is known 
(except for $\ci_6$; see below). They read
\ba
\ci_6 &\equiv& \ci(1,1,1,0,1,1,1,0,0,0)_{[6,952]} \;, \la{eq:I6} \\
\ci_8 &\equiv& \ci(1,1,1,1,1,0,0,0,0,1)_{[6,993]} \;, \\
\ci_{14} &\equiv& \ci(1,1,1,1,1,1,0,0,-1,1)_{[7,1009]} \;, \\
\ci_{15} &\equiv& \ci(1,1,1,2,1,1,1,0,0,-2)_{[7,1016]} \;, \\
\ci_{16} &\equiv& \ci(1,1,1,3,1,1,1,0,0,-3)_{[7,1016]} \;, \\
\ci_{17} &\equiv& \ci(1,1,1,1,1,1,-1,-1,1,1)_{[8,1011]} \;, \\
\ci_{18} &\equiv& \ci(1,1,1,1,1,1,0,-2,1,1)_{[8,1011]} \;, \\
\ci_{19} &\equiv& \ci(1,1,1,1,1,1,1,1,-2,0)_{[8,1020]} \;, \\
\ci_{20} &\equiv& \ci(1,1,1,1,1,1,1,1,0,-2)_{[8,1020]} \;, \\
\ci_{21} &\equiv& \ci(1,1,1,1,1,1,1,1,1,-3)_{[9,1022]} \;. \la{eq:I21}
\ea
Note that sectors [841], [1010] and the non-planar [511] are 
not present.\footnote{Remarkably, all non-planar vacuum Feynman diagrams in Yang-Mills theory at four 
loops vanish identically after performing the color trace over the totally antisymmetric structure constants $f^{abc}$.}
Within the family of 4-loop non-factorized sectors, we can identify three separate subcategories ${\cal S}_i$ based 
on the types of subgraphs these graphs are built from, suggesting in this way a hierarchy of complexity in their evaluation. 

First, the subset ${\cal S}_1=\{ \ci_6, \ci_{15}, \ci_{16} \}$ contains members of sectors [952] and [1016], 
made up of three simple 1-loop 2-point subgraphs. 
The machinery for tackling such types of vacuum graphs was developed in \cite{Arnold:1994ps}, 
relying on systematic subtractions of divergences present in the subgraphs, essentially mapping the problem to 
a numerical finite contribution plus a divergent lower-loop piece. 
Up to three loops, a number of sum-integrals (given in \app \ref{se:3loop}) have been evaluated using this method.
Similarly, the only 4-loop bosonic vacuum sum-integral that has ever been computed \cite{Gynther:2007bw} 
belongs to sector [952] and heavily relied on this approach as well.
Concretely, in \cite{Gynther:2007bw} a specific linear combination of sum-integrals had been evaluated,
see our \app\ref{se:4loop} for the detailed form. The motivation for considering that linear combination
had come from including terms that arise from lower loop orders when renormalizing the coupling constant
of the scalar theory considered in \cite{Gynther:2007bw}. In practice, this helped cancelling logarithmic 
ultraviolet divergences present in the 1-loop 2-point subgraphs.
In terms of the (UV-subtracted) 4-loop sum-integral $S_2$ evaluated in \cite{Gynther:2007bw}
and given in \app\ref{se:4loop}, we can at least rewrite
\ba
\ci_6 &=& \Lambda^{-8\ep}\,S_2 + \frac3\ep\Big[\lim_{\ep\to0}\ep\,I_2^0\Big]B_2 \;.
\ea
Here, $B_2$ is a 3-loop basketball-type sum-integral from \app\ref{se:3loop}, whose ${\cal O}(\ep)$ term
that would be required to expand $\ci_6$ up to the constant term is currently unknown, however.
The term in square brackets is finite; it extracts the coefficient of the prototype divergence of one-loop 
sum-integrals (see \app\ref{se:1loop}) and equals $(4\pi)^{-2}$.

The two sum-integrals within sector [1016] in ${\cal S}_1$, in contrast, contain non-trivial numerators. 
In analogy to the procedure carried out in \se \ref{se:eval3loop}, these can be mapped onto (partially) known 
structures upon tensor decomposition of 1-loop 2-point functions, 
slightly generalizing the 1-loop tadpole decomposition formulae of \cite{Navarrete:2022adz}.
We obtain
\ba
\ci_{15} &=& 12\,\tilde{\ci}_{15} - 3\,\ci_9 - \frac{1}{2}\,\ci_6 + 6\,\ci_5\;, \\
\ci_{16} &=& 48\,\tilde{\ci}_{16}-\frac{3}{2}\,\ci_{15} + 9\,\ci_{12} - 12\,\ci_{11} + 12\,\ci_{10} - 6\,\ci_9 
- \frac{1}{2}\,\ci_6 + 24\,\ci_5\;,
\ea
where on the right-hand side we encounter factorized 4-loop sum-integrals evaluated in \secs\ref{se:eval2loop} 
and \ref{se:eval3loop}, as well as two new 4-loop sum-integrals that contain explicit 1-loop tensor 2-point functions
\ba \la{eq:I15tilde}
&&\tilde{\ci}_{15} \;\equiv\; \Tint{K}\, \Pi(K)\;\Pi^{\mu\nu}(K)\,\Pi^{\mu\nu}(K) \;,\quad 
\tilde{\ci}_{16} \;\equiv\; \Tint{K}\, \Pi^{\mu\nu}(K)\,\Pi^{\nu\rho}(K)\,\Pi^{\rho\mu}(K)\;, \\
&&\mbox{with}\quad
\Pi(K) \equiv \Tint{Q} \frac1{Q^2\,(Q-K)^2} \;,\quad
\Pi^{\mu\nu}(K) \equiv \frac1{K^2}\,\Tint{Q} \frac{Q^{\mu}\,Q^{\nu}}{Q^2\,(Q-K)^2} \;,
\ea
and still need to be evaluated. To this end, we observe that a similar mapping had already been performed when 
evaluating the 3-loop sum-integrals $B_3$ and $J_{13}$ of \app\ref{se:3loop}, 
where tensor structures of this form were dealt with using a general decomposition method based on dimensional shifts, 
followed by the standard subtraction of divergences of the 1-loop subgraphs \cite{Ghisoiu:2012yk}. 
While one still expects such a strategy to be applicable in \eq\nr{eq:I15tilde}, 
the presence of an additional $\Pi$\/-insertion as compared to the 3-loop counterpart 
significantly increases the number of subtraction terms, 
in addition to the possible appearance of various divergences of the form
\ba
\frac{1}{\ep}\; \Tint{K} \frac{1}{[K^2]^{s}}\, \Pi(K) \Pi(K)
\ea
that had already affected the direct evaluation of $\ci_6$ discussed above. 
This could motivate the search for completely new evaluation methods altogether, going beyond the 
proven 1-loop subtraction strategies.

Second, the subset ${\cal S}_2=\{\ci_8,\ci_{14},\ci_{19},\ci_{20}\}$ has one member in sector [993], one in [1009] and two in [1020]. 
These consist of insertions of 1-loop and 2-loop 2-point subgraphs, the latter of which requires
a generalization of the 1-loop sub-divergence subtraction strategy.
Within this set, $\ci_{8}$ might be a nice starting point as it does not contain numerators.

Third, the final subset ${\cal S}_3=\{\ci_{17},\ci_{18},\ci_{21}\}$ consists of two sum-integrals falling into sector [1011] 
and one in [1022]. Their structure appears the most intricate of our list, 
being made up from two subgraphs with 3-point structure, for which there does not seem to 
exist previous experience regarding evaluation methods.

%
\section{Conclusions}
\label{se:conclu}

We have reported on major progress towards solving a long-standing open problem in finite-temperature field theory.
The last missing piece in an effective theory approach as sketched in \cite{Braaten:1995jr}
and applied to evade an infrared problem that concerns the QCD pressure
has been reduced into what now seems a manageable form.
To do so, we have confronted the open 4-loop strict perturbative expansion of the hard-mode contributions
to the pressure, summing up the 65 Feynman diagrams shown in \fig\ref{fig:4loop} and
manipulating the corresponding sum-integrals with modern algorithmic methods.

Bringing the integrand into canonical form has enabled us to reduce the number of Feynman sum-integrals
that contribute to the four-loop pressure of hot SU($\Nc$) gauge theory from order $10^5$ down to 21, at the
same time obtaining explicit gauge-parameter independence.
Furthermore, we have linked the evaluations of 11 of these 21 remaining masters to previously known results,
such that only 10 genuine 4-loop sum-integrals remain to be considered, as listed in \eqs\nr{eq:I6}--\nr{eq:I21}.
The evaluation of which, up to the constant term in the $\ep$-expansion, we leave as an exercise for the reader.

Once the ten remaining 4-loop sum-integrals are available,
it will be a most enjoyable task to collect all ingredients of the effective theory setup
and put them together for the long sought-after ``physical leading-order'' result
for the hot Yang-Mills pressure, and to present an update on issues such
as series convergence, (renormalization) scale dependence and a parameter-free comparison to lattice data.

Looking forward, a generalization of our canonical approach to include quarks 
is not difficult, as they do not self-interact and hence only occur in closed loops.
Therefore, one merely needs to keep track of the fermion signature of the four loop momenta $K_j$
when performing momentum shifts, and we expect similarly spectacular reductions for unintegrated 
expressions in full thermal QCD at four loops as those observed here for the $\Nf=0$ case.
Furthermore, the setup can then be easily extended to the case of dense QCD by keeping
track of signs of propagator momenta \cite{Navarrete:2024zgz}.

Finally, we anticipate that it might even be feasible to compute a correction to the full 4-loop pressure.
Indeed the next term in the effective field theory setup is of order $g^7$ and entails a five-loop computation
within 3-dimensional gauge theory coupled to a massive scalar (known as EQCD in the literature). 
Technically, this corresponds to regular (zero-temperature) continuum vacuum Feynman integrals
of QED-type, which coincidentally are currently being studied \cite{Maier:2024rng} in the collider physics community.

%
\acknowledgments

P.N.\ has been supported by an ANID grant Mag\'ister Nacional No.\ 22211544, 
by the Doctoral School of the University of Helsinki, 
and by the Academy of Finland grants No.\ 347499, 353772 and 1322507.
Y.S.\ acknowledges support from ANID under FONDECYT projects No.\ 1191073 and 1231056,
and from UBB/VRIP project No.\ EQ2351247.
All figures have been prepared with Axodraw \cite{Collins:2016aya}.

%
\appendix

%
\section{Bare pressure up to $g^4$}
\la{se:123loop}

As a further check on the canonicalization procedure developed and applied for reducing the four-loop term $p_4$
to its compact form \eq\nr{eq:p4can}, we can apply the same recipe for the two- and three-loop pressure.

We again start with two different strategies to generate the diagrams as explained in \se\ref{se:qgraf},
obtaining the sets shown in \figs\ref{fig:12loop} and \ref{fig:3loop}. 
Next, we re-run the same classification, scalarization and canonicalization algorithm of \se\ref{se:canonical}
over the sum-integrals corresponding to those diagrams.
Adding the $d$\/-dimensional form of the leading-order term (for a recent textbook see \cite{Laine:2016hma}), 
we obtain the coefficients of \eq\nr{eq:pHard} (at $\Nf=0$)
\ba
p_0(d,T) &=& -(\Nc^2-1) \frac{d-1}2\,\Tint{K} \ln(K^2) \;,\\
p_2(d,T) &=& -(\Nc^2-1) \frac{(d-1)^2}4\,\Tint{K}\,\Tint{Q}\, \frac1{K^2\;Q^2} \;,\\
p_3(d,T) &=& (\Nc^2-1) \frac{(d-1)^2}8\,\Big[ 2(d-5)\,\ci(2,1,1,0,0,0) + \ci(1,1,0,0,1,1) 
\nonumber\\&&\hphantom{(\Nc^2-1) \frac{(d-1)^2}8\,\Big[}
-4\,\ci(1,1,1,1,0,0) + 2\,\ci(2,1,1,1,1,-2)\Big] \;,
\ea
where for the 3-loop pressure our algorithm has employed the corresponding 6-element momentum 
family \eq\nr{eq:family3loop}.

Finally, we follow the strategy of \se\ref{se:evals} and identify factorizable sectors, which in this case do
not have numerators and decompose trivially into 1-loop factors $I_\nu^\eta$ of \app\ref{se:1loop}. 
For the remaining sum-integrals in $p_3$ we use the naming scheme of \app\ref{se:2loop} and \ref{se:3loop}
which results in
\ba
p_0(d,T) &=& -(\Nc^2-1) \frac{d-1}d\,I_1^2 \;,\\
p_2(d,T) &=& -(\Nc^2-1) \frac{(d-1)^2}4\,(I_1^0)^2 \;,\\
p_3(d,T) &=& (\Nc^2-1) \frac{(d-1)^2}8\,\Big[ 2(d-5)\,I_2^0\;(I_1^0)^2 + B_2 -4\,\sumintL_{1,1,1}^{0,0,0}\;I_1^0 + 2\,B_3 \Big] \;.
\ea 
Note that the two-loop sum-integral $\sumintL_{1,1,1}^{0,0,0}=0$ (cf.\ \eq\nr{eq:L111}) could be dropped from the last line.
These expressions coincide with the known results, see e.g.\ \cite{Braaten:1995jr,Nishimura:2012ee},
and should come in handy when passing from the bare to the renormalized pressure.

%
\section{Massless bosonic vacuum sum-integrals}
\label{se:scalar}

Here, we collect known results for scalar massless bosonic vacuum sum-integrals up to the four-loop level. 
Starting at two loops, we refrain from displaying the explicit expressions,
but rather point to the original references where those results can be found.

One technical detail that was not needed in the main text but is relevant here is the following.
The presence of a heat bath breaks Lorentz invariance down to just rotations in three-dimensional Euclidean space, 
introducing a four-vector $U=(1,\vec{0})$ on which tensor structures may depend. 
As a consequence, powers of $U^\mu K^\mu \equiv U\ccdot K = K_0$ can appear in the numerators of sum-integrals.
We keep track of such numerator powers wherever required by simply adding four more list entries 
in \eq\nr{eq:family} (three in \eq\nr{eq:family3loop}).
All techniques explained in the main text, such as sector classification, momentum shifts, 
or canonicalization of the sum-integrals apply to such augmented index lists as well.

%
\subsection{One loop}
\la{se:1loop}

The most general scalar 1-loop massless bosonic vacuum sum-integral can be computed analytically 
in $d$ dimensions ($\eta\in\setN_0$ and $\nu\in\setZ$):
\ba \la{eq:I}
I_\nu^\eta(d,T) &\equiv& \Tint{K} \frac{(K_0)^\eta}{[K^2]^\nu} 
\;=\;  \frac{[1+(-1)^\eta]\,T\,\zeta(2\nu-\eta-d)}{(2\pi T)^{2\nu-\eta-d}}\,
\frac{\Gamma(\nu-\tfrac{d}2)}{(4\pi)^{d/2}\,\Gamma(\nu)} \;.
\ea
Note that $I_\nu^\eta$ vanishes for odd values of $\eta$ as a consequence of the integrand's symmetry 
under $K_0\to-K_0$. 

Another ``one-loop'' case of interest is the type of sum-integral that enters the ideal gas pressure, 
arising from the logarithms of a Gaussian path integral over the quadratic part of the action.
From dimensional analysis and taking a derivative with respect to the temperature, 
it can be related \cite{Nishimura:2012ee} to the integrals of \eq\nr{eq:I} as
\ba
\Tint{K} \ln(K^2) &=& \frac2{d}\,I_1^2(d,T) \stackrel{d=3-2\ep}{\approx} -\frac{\pi^2 T^4}{45} + {\cal O}(\ep)\;.
\ea

%
\subsection{Two loops}
\la{se:2loop}

The most general scalar 2-loop massless bosonic vacuum sum-integral has the form
\ba\la{eq:L}
\sumintL_\nabc^\eabc(d,T) &\equiv& 
\Tint{K}\,\Tint{Q}\, \frac{(K_0)^{\eta_1}\,(Q_0)^{\eta_2}\,(K_0-Q_0)^{\eta_3}}{[K^2]^{\nu_1}\,[Q^2]^{\nu_2}\,[(K-Q)^2]^{\nu_3}} \;.
\ea
It turns out that for any combination of indices $\eta_i\in\setN_0$ and $\nu_i\in\setZ$, $\sumintL_\nabc^\eabc$ 
reduces to a sum of products of two 1-loop tadpoles $I_\nu^\eta$, as given in \eq\nr{eq:I}. 
This reduction is a consequence of integration-by-parts (IBP) identities that can either be applied
at fixed integer values of $\{\eta_i,\nu_j\}$ \cite{Nishimura:2012ee}, 
or even solved for the general case \cite{Davydychev:2022dcw}, leading to an explicit closed formula for the reduction
$\sumintL \rightarrow I\times I$ in the general-index case \cite{Davydychev:2023jto}.

%
\subsection{Three loops}
\la{se:3loop}

At the 3-loop level, all available results have been obtained only up to the constant term in the $\ep$ expansion 
in dimensional regularization, making use of the methods pioneered by Arnold and Zhai \cite{Arnold:1994ps}.
In the following, we employ the 3-loop propagator momentum family $\{P_a\}$ defined in \eq\nr{eq:family3loop}
and add three additional indices to our index-list to account for factors of Matsubara frequencies in the numerators.
The general vacuum sum-integral can then be represented in our list-notation as (recall $U=(1,\vec{0})$)
\ba \la{eq:3looplist}
\Tint{K_1}\Tint{K_2}\Tint{K_3} 
\frac{(U\ccdot K_1)^{m_1}(U\ccdot K_2)^{m_2}(U\ccdot K_3)^{m_3}}
{(P_1\ccdot P_1)^{s_1}\dots (P_6\ccdot P_6)^{s_6}} 
\equiv \ci(s_1,\dots,s_6;m_1,\dots,m_3) \;.
\ea

At mass dimension 4 (relevant the 3-loop pressure), the following sum-integrals are known 
(see {\cite{Nishimura:2012ee} for an overview)
\ba
B_2 &=& \ci(1,1,0,0,1,1;0,0,0) \quad 
\mbox{($I_\rmi{ball}$ in {\cite{Arnold:1994ps}}; {${\cal M}_{0,0}$ in \cite{Schroder:2012hm}})} \;,\\
B_3 &=& \ci(2,1,1,1,1,-2;0,0,0) \quad 
\mbox{(related to $I_\rmi{sqed}$ in {\cite{Arnold:1994ps}}; {${\cal M}_{2,-2}$ in \cite{Schroder:2012hm}})} \;.
\ea

At mass dimension 2 (relevant for the matching parameter $m_\rmi{E}^2$ in EQCD), 
the following sum-integrals are known (see \cite{Ghisoiu:2015uza} for an overview)
\ba
J_{11} &=& \ci(1,1,1,1,1,0;0,0,0) \quad \mbox{($I$ in \cite{Andersen:2008bz}; ${\cal M}_{1,0}$ in \cite{Schroder:2012hm})} \la{eq:J11} \;,\\
J_{12} &=& \ci(2,1,1,1,1,0;0,2,0) \quad \mbox{($V_2$ in \cite{Ghisoiu:2012kn})} \;,\\
J_{13} &=& \ci(3,1,1,1,1,-2;0,0,0) \quad \mbox{(${\cal M}_{3,-2}$ in \cite{Ghisoiu:2012yk})} \la{eq:J13} \;,\\
J_1 &=& \ci(2,1,0,0,1,1;0,0,0) \quad \mbox{($S_1$ in \cite{Gynther:2007bw}; $B_2$ in \cite{Moller:2010xw})} \la{eq:J1} \;,\\
J_4 &=& \ci(3,1,0,0,1,1;2,0,0) \quad \mbox{($B_{3,2}$ in \cite{Moller:2012chx})} \;.
\ea
We note that these integrals are not all independent, since one can derive 
a linear relation between three of them via IBP methods (see \app B of \cite{Moller:2012chx}),
\ba
0=3(d-3)^2(d-4)\,J_{11} -2(3d^2-24d+47)\,J_1 -16(d-4)\,J_4 \;.
\ea

At mass dimension 0 (relevant for the effective coupling $g_\rmi{E}^2$ in EQCD), 
the following sum-integrals are known (see \cite{Ghisoiu:2013zoj} for an overview)
\ba
V_1 &=& \ci(1,2,1,1,1,0;0,0,0) \quad \mbox{($V_1$ in \cite{Ghisoiu:2012ph})} \;,\\
V_2 &=& \ci(2,1,1,1,1,0;0,0,0) \quad \mbox{($V(3;21111;000)$ in \cite{Ghisoiu:2013zoj})} \;,\\
V_3 &=& \ci(2,2,1,1,1,0;0,0,2) \quad \mbox{($V(3;22111;002)$ in \cite{Ghisoiu:2013zoj})} \;,\\
V_4 &=& \ci(3,1,1,1,1,0;0,2,0) \quad \mbox{($V(3;31111;020)$ in \cite{Ghisoiu:2013zoj})} \;,\\
V_5 &=& \ci(4,1,1,1,1,0;0,2,2) \quad \mbox{($V(3;41111;022)$ in \cite{Ghisoiu:2013zoj})} \;,\\
V_6 &=& \ci(3,1,0,0,1,1;0,0,0) \quad \mbox{($B_3$ in \cite{Moller:2010xw})} \;.
\ea
Of these, only $V_1$, $V_2$ and $V_4$ are needed for our factorizations of \se\ref{se:eval3loop}.

%
\subsection{Four loops}
\la{se:4loop}

To date, a single genuine (ultraviolet-subtracted, bosonic) 4-loop sum-integral has been evaluated. 
It is the only sum-integral at the 4-loop level contributing to the (renormalized) pressure of massless scalar $\phi^4$ theory. 
Using the scalar 1-loop two-point sum-integral
\ba \la{eq:Piscalar}
\Pi(K)= \Tint{Q} \frac{1}{Q^2(K-Q)^2}\;,
\ea
the 4-loop sum-integral that has been evaluated in \cite{Gynther:2007bw} reads
\ba\label{eq:S2}
S_2 \equiv \Lambda^{\!8\ep}\,\Tint{K} \Big( [\Pi(K)]^3 - \frac{3\,\Lambda^{-2\ep}}{(4\pi)^2\ep}\,[\Pi(K)]^2 \Big) 
\;\approx\; -\frac{T^4}{(4\pi)^4}\frac{1}{16\ep^2}\,\Big( 1+\ep\,t_{11} + \ep^2\,t_{12} \Big) 
+ {\cal O}(\ep)\,. \qquad
\ea
Here, $t_{11}$ is a fully analytic coefficient and $t_{12}$ is numerical. The subtraction term appearing 
in \eq\nr{eq:S2} originates from gauge coupling renormalization in the above-mentioned theory. 
The inclusion of this counterterm significantly simplifies the computation of this sum-integral, 
as one avoids the need to evaluate a three-loop sum-integral up to $O(\ep)$, a feat that goes beyond 
current sum-integral technology.

%

%
\end{document}